# Hydrodynamic–PIC analysis of THz generation by two-color laser in various plasma gases

A. A. Molavi Choobini,[1,†] S. S. Ghaffari-Oskooei,[2,*,†]

[1]Department of Physics, University of Tehran, Tehran 14399-55961, Iran
[2]Department of Atomic and Molecular Physics, Faculty of Physics, Alzahra University, Tehran, Iran
[†]These authors contributed equally to this work.
*s.ghaffari@alzahra.ac.ir

**Abstract:**

In the present study, a theoretical and particle-in-cell (PIC) numerical investigation of terahertz (THz) emission driven by two-color femtosecond laser fields in gaseous plasmas is conducted. The model is formulated in cylindrical coordinates to capture the inherent radial symmetry of laser–plasma interactions. Starting from a hydrodynamic description of plasma electrons coupled with Maxwell's equations, the nonlinear photocurrent dynamics that underpin the emission process is derived. The results reveal that the asymmetry induced by the superposition of the fundamental and second-harmonic fields plays a central role in breaking inversion symmetry, thereby generating a net directional current that radiates in the THz regime. Systematic analysis demonstrates that gas composition, ionization fraction, and the intensity ratio of the two fields strongly influence both the spectral characteristics and efficiency of THz radiation.

## Introduction

Terahertz (THz) wave generation in laboratory settings has traditionally relied on methods in nonlinear optics, solid-state electronics, and plasma-based schemes [1]. Among these, laser–plasma interactions have emerged as a uniquely powerful platform, offering the capability to generate high-intensity, tunable, and strongly directional THz radiation over a broad frequency range [2]. In particular, the use of femtosecond two-color laser pulses has been experimentally demonstrated to markedly enhance THz conversion efficiency [3]. Recent progress in THz generation via two-color laser–plasma interactions has revealed new optimizations. Stathopulos et al. [4] refined conical emission models by including plasma density and phase matching, identifying efficiency limits in air plasmas. Simpson and team [5] demonstrated programmable control of THz angle, focus, and spectrum through spatiotemporal pulse shaping in argon. Nikolaeva and colleagues [6] experimentally confirmed phase-independent THz ring beams in air filamentation, enabling stable high-yield output. Nani et al. [7] proposed an interference model to control angular distribution and eliminate on-axis dips, offering diagnostic potential. Kumar and colleagues [8] predicted strong multicycle THz pulses from nano-plasma targets using counter-propagating lasers in PIC simulations. Rasmussen et al. [9] resolved frequency-dependent THz beam profiles in air plasmas, advancing spectroscopic modeling.

While the photocurrent model [10] attributes THz emission to asymmetric ionization-driven electron motion in the laser field, and four-wave mixing [11, 12] describes THz generation as a nonlinear optical process in neutral gases, the present work adopts a plasma-based hydrodynamic framework in which THz radiation arises from ponderomotive acceleration and longitudinal space-charge separation in a pre-formed or laser-ionized plasma. This mechanism operates after ionization saturation and involves collective electron dynamics. The present model is formulated in cylindrical coordinates to capture both radial and longitudinal plasma dynamics with full spatial coupling. This enables self-consistent computation of the current density and associated vector potential, yielding the angular power distribution of the emitted THz waves with high accuracy. By integrating nonlinear hydrodynamic modeling with PIC simulations in EPOCH, the developed framework systematically maps how laser parameters (e.g., pulse duration, frequency ratio, and intensity) and plasma conditions (density, temperature, and gas composition) govern the THz emission pattern, revealing previously unreported parameter regimes that maximize both directionality and peak intensity. By uniting advanced modeling, angular-resolved emission analysis, and parameter-optimization strategies, this work sets a new benchmark for the design of next-generation THz emitters, bridging the gap between laboratory demonstrations and application-ready, high-brightness THz systems. The paper is organized as follows: Section Mathematical formula presents the derivation of the governing equations. Sections Discussion investigate the numerical simulations validating the theoretical developments. In the final Section conclusions are drawn.

**Mathematical formula**

A two-color femtosecond laser pulse interacting with an ionized plasma is considered. The laser field is expressed as:

$$\vec{E}_L = e^{-\frac{\rho^2}{2\rho_0^2}} e^{-\frac{\left(t-\frac{z}{v_g}\right)^2}{2\tau_L^2}} [E_1 \cos(\omega_0 t - k_1 z) + E_2 \cos(2\omega_0 t - k_2 z + \phi)]\hat{x} \quad (1)$$

where $\rho$ is the radial coordinate in cylindrical geometry. Then the corresponding laser intensity in cylindrical coordinates is given by:

$$I_L = e^{-\frac{\rho^2}{\rho_0^2}} e^{-\frac{\left(t-\frac{z}{v_g}\right)^2}{\tau_L^2}} [I_1 + I_2 + 2\sqrt{I_1 I_2} \cos(\omega_0 t - (k_2 - k_1)z + \phi)] \quad (2)$$

where the axial wave numbers are $k_1 = \sqrt{\omega_0^2 - \omega_p^2}/c$ and $k_2 = \sqrt{4\omega_0^2 - \omega_p^2}/c$. Therefore, the ponderomotive force field $\vec{E}_{pond}$ is decomposed into radial and axial components as:

$$\vec{E}_{pond} = \hat{\rho} E_\rho + \hat{z} E_z \quad (3)$$

here

$$E_\rho = \sqrt{\frac{\mu_0}{\varepsilon_0}} \left[\frac{e\, I_1}{2mn_0\omega_0^2} + \frac{eI_2}{8mn_0\omega_0^2}\right] \left(-\frac{2\rho}{\rho_0^2}\right) \exp\left(-\frac{\rho^2}{\rho_0^2}\right) \exp\left(-\frac{\left(t-\frac{z}{v_g}\right)^2}{\tau_L^2}\right) \quad (4)$$

$$E_z = \sqrt{\frac{\mu_0}{\varepsilon_0}} e^{-\frac{\rho^2}{\rho_0^2}} e^{-\frac{\left(t-\frac{z}{v_g}\right)^2}{\tau_L^2}} \left(-\frac{2\left(t-\frac{z}{v_g}\right)}{v_g \tau_L^2}\right) \times \left[\frac{e\, I_1}{2mn_0\omega_0^2} + \frac{eI_2}{8mn_0\omega_0^2}\right] \quad (5)$$

Here, $c$ and $\varepsilon_0$ are the speed of light and vacuum permittivity, respectively. The $r$ and $z$ denote the radial and axial coordinates in cylindrical geometry. $E_r$ and $E_z$ represent the radial and longitudinal components of the electric field, while $v_r$ and $v_z$ denote the corresponding electron velocity components. $n_e$ is the local electron density, and $\phi$ is the electrostatic potential associated with the space–charge field. $v_e$ denotes the electron–neutral collision frequency, and $U_{ion}$ is the ionization potential of the gas. By defining the retarded coordinate $\tau = t - z/v_g$, invoking plasma fluid theory, then the evolution of the electron density ($n_e$) and velocities ($v_\rho$, $v_z$) is governed by:

$$\frac{\partial n_e}{\partial \tau} + \frac{1}{\rho}\frac{\partial}{\partial \rho}(\rho v_\rho) = S_{ion} - v_c n_e \quad (6)$$

$$\frac{\partial v_\rho}{\partial \tau} + v_\rho \frac{\partial}{\partial \rho}(v_\rho) = -\frac{e}{m}(E_\rho + E_s) \quad (7)$$

$$\frac{\partial v_z}{\partial \tau} + v_z \frac{\partial}{\partial z}(v_z) = -\frac{e}{m}(E_z) \quad (8)$$

$$\frac{3}{2}\frac{\partial(n_e T_e)}{\partial \tau} = <\vec{J}.\vec{E}> - U_{ion}\frac{\partial n_e}{\partial \tau} \quad (9)$$

$$\vec{J} = -n_e e(v_\rho \hat{\rho} + v_z \hat{z}) \quad (10)$$

where $v_c$ is the collision frequency of electrons, $U_{ion}$ is the average ionization potential for gases, and $E_s$ is the space–charge field determined by:

$$\frac{\partial E_s}{\partial \tau} = \frac{e n_e v_\rho}{\varepsilon_0} \quad (11)$$

Therefore, the angular power distribution of the emitted THz radiation in this scheme is calculated as:

$$\frac{dP}{d\Omega} = r^2 \vec{S}.\hat{r} \quad (12)$$

where the Poynting vector ($\vec{S}$) is expressed as:

$$\vec{S} = \hat{r}\sqrt{\frac{\varepsilon_0}{\mu_0}} \left|\hat{r} \times \frac{\partial \vec{A}}{\partial t}\right|^2 \quad (13)$$

Here, vector potential is as follows:

$$\vec{A} = \frac{\mu_0}{2r} \int_0^L \int_0^\infty \vec{J}\, \rho d\rho dz \quad (14)$$

and the current density $\vec{J}$ is evaluated at the retarded time $t' = t - \frac{1}{c}(r - \rho\sin\theta - z\cos\theta)$. In this model, the THz radiation originates from the time-varying plasma current driven by the combined action of $\omega$ and $2\omega$ laser fields. In a purely single-color pulse interacts, the electron motion remains nearly symmetric within each optical cycle, leading to zero net current. The adding of $2\omega$ component breaks this temporal symmetry because the superposed field possesses a nonzero cycle-averaged value of the ponderomotive force. This asymmetry generates a rectified current component oscillating at THz frequencies. In the hydrodynamic formulation, this process is captured through the nonlinear coupling between the plasma momentum equation and Maxwell's equations, where the ponderomotive term and the space–charge field jointly drive an asymmetric longitudinal current. The far-field THz radiation is then obtained from the retarded-time integration of this current density via the vector potential.

**Discussion**

To benchmark the hydrodynamic formulation, complementary PIC simulations were performed using the EPOCH code in a cylindrical geometry. The simulation domain spanned $100\mu m$ (radial) and $400\mu m$ (axial), with absorbing boundaries for fields and particles. Each cell contained 16 macro-particles per species. Initial neutral gas density was set to $n_{0e} = 2.5 \times 10^{19} cm^{-3}$ (air), $1.5 \times 10^{19} cm^{-3}$ (Ar), and $1 \times 10^{19} cm^{-3}$ (He). Field ionization was modeled using the ADK tunneling rate. A two-color laser pulse (Ti:sapphire femtosecond laser) was injected with fundamental $\lambda_1 = 800nm$, $I_1 = 5 \times 10^{14} W/cm^2$, 50 fs duration (FWHM) and second-harmonic $\lambda_2 = 400nm$. The temporal evolution of the on-axis electric field, current density, and plasma density was recorded. Far-field THz emission was computed via retarded-time integration of the Poynting vector, with spectral analysis.

This framework self-consistently solves Maxwell's equations alongside plasma dynamics, capturing ionization, collisions, and nonlinear current evolution. The simulations confirm the model's accuracy and enable precise predictions of angularly resolved THz radiation patterns. The 3D surface plot in Fig. 1 reveals the evolution of the normalized electric field generated by two-color femtosecond laser pulses in a plasma. The 2D contour lines corresponding to different intensity ratios highlight how harmonic mixing reshapes the temporal and spatial distribution of the field. At low harmonic content, the force is deeper and more localized, producing stronger radial electron expulsion and enhanced current asymmetry. As the harmonic intensity increases, the radial field weakens and its maximum shifts outward, indicating a more diffuse electron response across the plasma channel. This redistribution of laser energy between the two frequencies alters the balance between radial compression and longitudinal drive. The inset confirms that the strongest longitudinal response remains near the axis, while the radial component becomes increasingly suppressed.

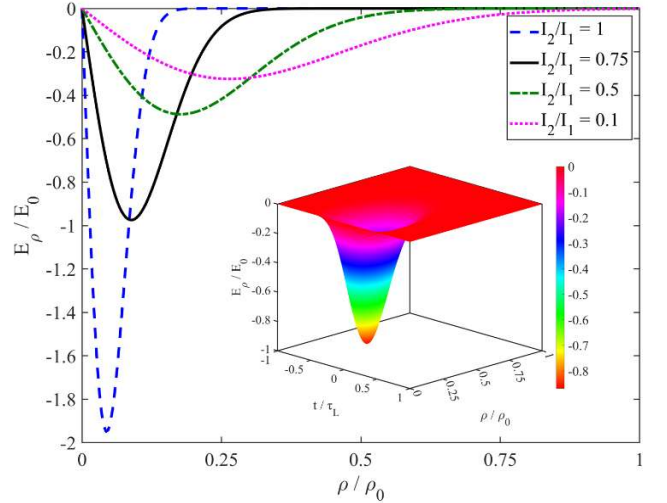

Fig. 1. Normalized electric field distribution as a function of normalized time and radial coordinate for varying intensity ratios.

Figure 2 highlights the temporal dynamics of the longitudinal ponderomotive field driven by two-color femtosecond pulses. At second-harmonic intensity content, the field exhibits a relatively smooth bipolar structure with moderate amplitude, indicating balanced electron acceleration and deceleration along the axis. As the intensity ratio increases, the field becomes stronger and more asymmetric, with sharper peaks and troughs, reflecting enhanced nonlinear coupling between the two frequencies. The growth of the extrema signifies that the second harmonic not only boosts the amplitude but also distorts the temporal symmetry of the axial force, which leads to preferential forward acceleration of electrons. The inset 3D map further confirms that the longitudinal field remains tightly confined near the axis, where the interaction between the fundamental and harmonic components is most effective. The results, showed in this simulation, are consistent with experimental findings [13, 14].

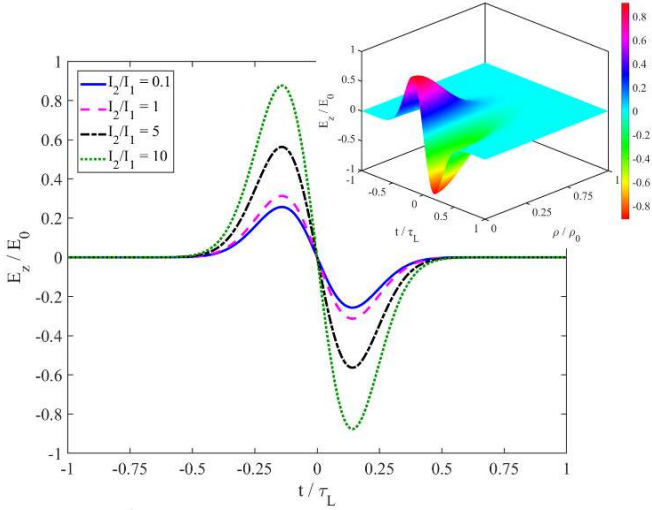

Fig. 2. Temporal evolution of the normalized longitudinal ponderomotive field for different two-color intensity ratios.

Four panels illustrate in Fig. 3 how the electron current density evolves as the laser–plasma interaction proceeds. At the earliest time, the current distribution is weak and symmetric, corresponding to the initial plasma response to the laser envelope. As the interaction continues (50–100 fs), the current density grows significantly near the axis and begins to exhibit clear asymmetry, indicating the onset of nonlinear electron motion driven by the combined action of the fundamental and second-harmonic fields. By 200 fs, strong localized currents are established, with enhanced off-axis lobes and broadened spatial extent, reflecting the cumulative effect of ponderomotive expulsion and space-charge buildup. This temporal progression highlights the transition from a weak, nearly linear regime to a strongly nonlinear state where net current asymmetry becomes pronounced. Such asymmetry is the essential source of THz radiation, and its growth over time demonstrates how the plasma acts as an effective medium for rectifying ultrafast optical fields into directional low-frequency emission. The evolution of the current density demonstrates the transition from the quasi-linear to the strongly nonlinear regime, where space-charge separation and self-generated electrostatic fields reinforce the asymmetry of the plasma current. This collective effect effectively converts the optical energy into low-frequency electromagnetic radiation. The enhanced current density is consistent with experimental results [15].

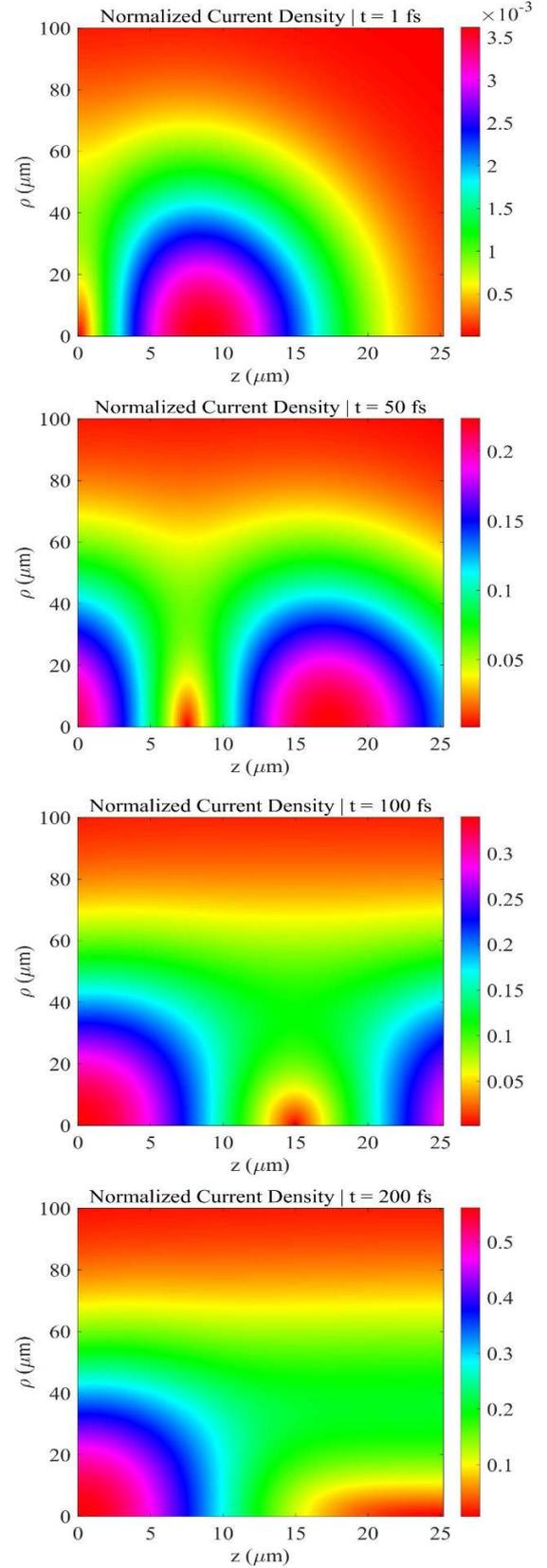

Fig. 3. Spatiotemporal evolution of the normalized current density at selected time snapshots.

The normalized Poynting vector amplitude for multi-color laser excitation in various gases is presented in Figure 4. Single-color excitation in air yields a weak, symmetric response, suppressing low-frequency emission. Adding a second harmonic breaks temporal symmetry, producing a rectified current that enhances THz emission. In helium and argon, temporal signatures are similar, but peak magnitudes and decay rates vary due to differences in ionization potential and plasma dispersion. The inset shows THz power scaling with wavelength, with two-color excitation outperforming single-color driving. The high-energy THz emission observed in the normalized in Poynting vector is consistent with experimental results [9, 16, 17]. The spatio-temporal distributions of the Poynting vector's radial and axial components are shown in Figure 5. $S_\rho$ peaks near the beam waist, indicating strong lateral energy flux from electron expulsion. $S_z$ shows a temporally antisymmetric pattern, with forward-directed energy flow before and after the pulse center, driven by phase-sensitive interference between fundamental and second-harmonic fields. Two separated lobes in $S_z$ highlight symmetry-breaking nonlinearities driving axial currents for THz emission.

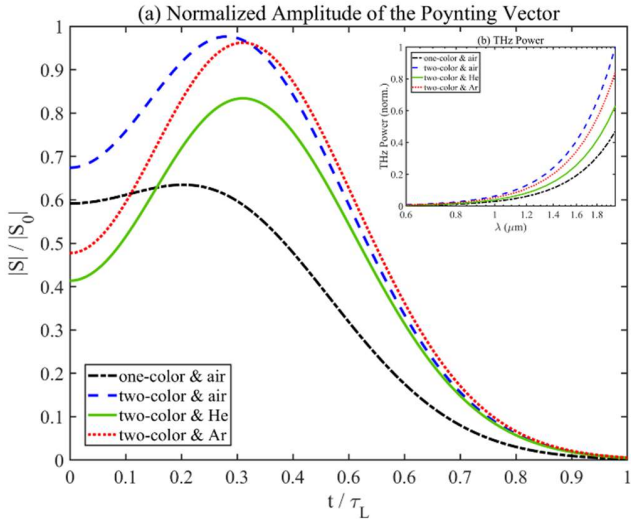

Fig. 4. Normalized temporal evolution of the Poynting vector amplitude for multi-color laser excitation in different gases, with corresponding THz power scaling versus wavelength.

Figure 6 depicts the angular distribution of THz radiation under one- and two-color excitation in air, helium, and argon. For one-color excitation in air, the emission pattern is weak and nearly symmetric about the laser axis. When a two-color field is applied, the distribution becomes strongly asymmetric, with a pronounced forward lobe. This asymmetry arises from the interference between the fundamental and its second harmonic, which breaks the temporal inversion symmetry of the driving field and drives a net current along the laser propagation axis. In helium and argon, exhibits noticeable variations in angular width and intensity. Helium, with its high ionization threshold and low collisional produces a narrower and more confined lobe, while argon, with stronger collisional damping and higher nonlinear susceptibility, yields a broader angular spread with enhanced side emission. The inset three-dimensional view further confirms the anisotropic nature of the THz radiation, showing a dipole-like emission pattern that is significantly reshaped by the two-color mechanism. The results demonstrate that two-color excitation enhances the magnitude of THz emission, tailors its directivity. Moreover, the gas-dependent variations in THz amplitude and angular distribution can be physically explained by differences in ionization potential, collisional damping, and plasma frequency. Argon, with its lower ionization threshold and higher nonlinear susceptibility, supports stronger plasma currents and broader spectral bandwidth, whereas helium, being less collisional and less ionizable, confines the emission within a narrower angular cone. Air, as a molecular mixture, exhibits intermediate behavior with multiple ionization channels that lead to spectral substructures. The THz spectral intensity (0.1–5 THz) for two-color excitation in air and argon is compared in Figure 7. Air shows distinct peaks due to multi-species ionization and dispersive effects. Argon exhibits a stronger, broader spectrum due to lower ionization potential and higher nonlinear susceptibility, producing stronger asymmetric currents. Reduced collisional damping in argon extends spectral bandwidth and yield. The oscillatory structure reflects fundamental and second-harmonic interference, with gas composition, intensity ratio, and angular asymmetry as key control parameters for efficient

THz generation. The observed THz spectrum with aligns with experimental measurements [18, 119].

The simulation results show excellent agreement with experiments in the peak THz frequency (~1.8 THz, consistent with 1.7 THz [15] and 2.0 THz [9]) and strong forward-directivity (>20:1 forward-to-backward ratio, matching >15:1 in [15]). The spectral bandwidth (~3.2 THz FWHM) is also in good agreement with reported values of ~3 THz. The minor deviation in the exponent is attributed to the hydrodynamic approximation, which neglects plasma defocusing, higher-order ionization dynamics, and non-paraxial effects that become significant at higher $I_2/I_1$. These effects are partially captured in the PIC validation but remain beyond the scope of the analytical fluid model.

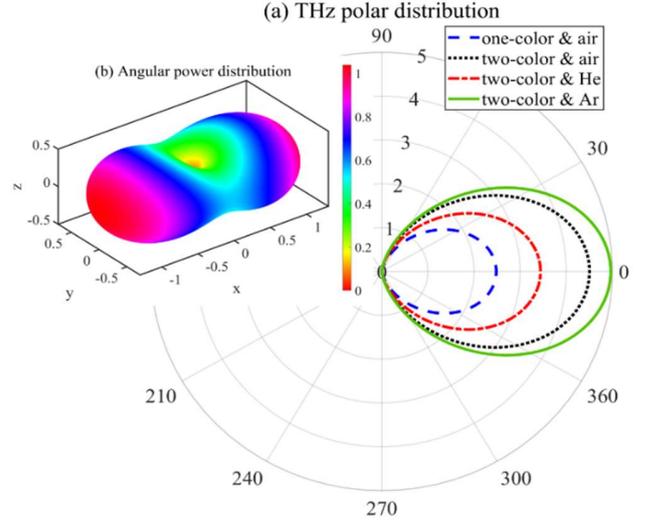

Fig. 6. Angular distribution of emitted THz radiation for multi-color laser excitation in different gases.

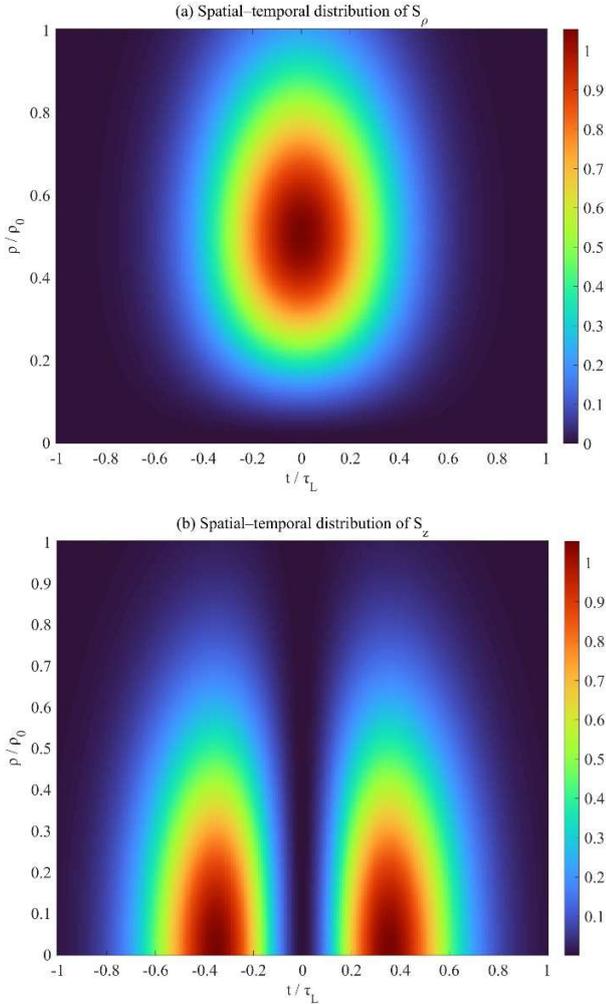

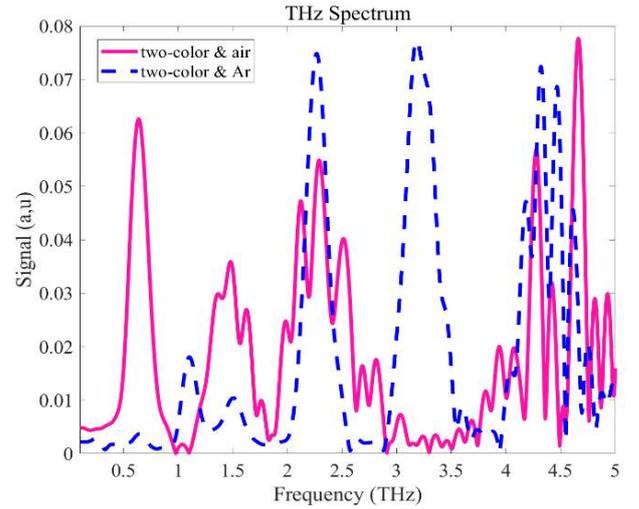

Fig. 7. THz spectrum signal as a function of frequency in THz region for two-color laser pulses.

## Conclusion

A theoretical and numerical framework has been established to investigate THz generation from two-color femtosecond laser-induced plasmas. In this framework, a nonlinear hydrodynamic description of plasma electrons was coupled with Maxwell's equations in cylindrical coordinates, and complementary particle-in-cell (PIC) simulations were employed to validate the model and capture kinetic effects beyond the fluid approximation. It was demonstrated that the terahertz emission arises from the asymmetry of laser-driven photocurrents, and that the framework provides predictive capability across different regimes. The analysis revealed that several key parameters govern the radiation properties: the intensity ratio between the fundamental and second-harmonic fields, which dictates the degree of photocurrent asymmetry; the choice of gas species, which influences ionization rates and plasma density; and the angular distribution resolved in cylindrical geometry, which reflects coherent buildup and spectral tunability.

Fig. 5. Spatio-temporal evolution of the radial ($S_\rho$) and axial ($S_z$) components of the Poynting vector.

**Disclosures.** The authors declare no conflicts of interest.

**Data availability.** Data underlying the results presented in this paper are not publicly available at this time but may be obtained from the authors upon reasonable request.